\documentclass[journal]{IEEEtran}
\usepackage{mathtools}
\usepackage{graphicx}
\usepackage{algorithm}
\usepackage{algorithmic}
\DeclareMathOperator*{\argmin}{argmin}
\usepackage{flushend}

\ifCLASSINFOpdf
\else
\fi
\usepackage{url}

\begin{document}
%
\title{Reinforcement Learning Based Algorithm for the Maximization of EV Charging Station Revenue}
%
%
%
\author{Stoyan~Dimitrov,~\IEEEmembership{PHD student, Sofia University}
        Redouane~Lguensat,~\IEEEmembership{Student, Telecom Bretagne}}

\maketitle

\begin{abstract}
This paper presents an online reinforcement learning based application which increases the revenue of one particular electric vehicles (EV) station, connected to a renewable source of energy. Moreover, the proposed application adapts to changes in the trends of the station's average number of customers and their types. Most of the parameters in the model are simulated stochastically and the algorithm used is the Q-learning algorithm. A computer simulation was implemented which demonstrates and confirms the utility of the model.
\end{abstract}

\begin{IEEEkeywords}
Reinforcement learning, electric vehicles, charging stations, renewable energy, Q-learning
\end{IEEEkeywords}

%
\IEEEpeerreviewmaketitle

\section{Introduction}
%
%
%
%
\IEEEPARstart{O}{ne} of the problems that hinders widespread use of electric vehicles (EV) is "range anxiety", users fear that their vehicle will stop in the middle of nowhere, because of the low battery level. The batteries of these plug-in hybrid electric vehicles are to be charged at home from a standard outlet or on a corporate car park. These extra electrical loads have an impact on the distribution grid such as power losses and voltage deviations. A mass domestic charging means that the vehicles are charged instantenously when they are plugged in or after a fixed start delay. This uncoordinated power consumption on a local scale can lead to grid problems\cite{Impact}.\\ 
\indent This fact explains increasing concerns with charging stations and it has led many researchers to work on theoretical models of the functioning of these stations. In our case, we present a model of an EV charging station, together with an allocation mechanism for its resources. The aim is to maximize station's revenue taking into consideration fluctuations of electricity price during the day.\\
\indent We named our algorithm SAEDS (Station Automated Electricity Distribution System). It is based on the area of machine learning called reinforcement learning. The goal of SAEDS is to learn how good are the different decisions of the station in the different states that it could be. This is performed in the so-called "learning phase". After the end of this phase, the application uses the already learned information to distribute electricity supply among the plugged-in vehicles.\\
\indent EV charging stations are connected to the electrical grid, which often means that their electricity comes from fossil-fuel power stations or nuclear power plants. In our model, it is assumed that the EV station is connected to a renewable source of energy. The fact, that the amount of electricity received from this source needs prediction, makes the model more complex. The complexity of the cosidered problem justifies the use of Machine Learning algorithm, instead of alternave approaches.
Sections II and III speaks about the literature we have used and about the theory behind the Q-learning method, respectively. Sections IV contains the details of the model we consider and the algorithm we use to solve it. Section V describes the results of the computer simulation, created to test SAEDS. The conclusion is stated in section VI.
\section{Related work}

 This work is very similar to those of O'Neill et al. in \cite{A}. There, they present an online learning application for residential demand response that uses Q-learning to learn the behavior of the consumers and to make optimal energy consumption decisions. We noticed that the same approach can be applied in the context of electric vehicles charging, viewing on the charging station as one big household, which wants to minimize its expenses, or equivalently, to maximize its incomes. The clients of the station, with their vehicles, play the same role as the electrical devices of the household in \cite{A}. Whether a vehicle or a household device will be plugged in at a particular moment is determined probabilistically in both papers.
 \par Another related publication (Valogianni et al. [6]) proposes a smart charging algorithm that uses Q-learning trained on real world data to learn the individual household consumption behavior. Additionally, an EV charging algorithm is proposed which maximizes individual welfare and reduces the individual energy expenses.\\
\indent Making the best energy consumption decisions was the purpose of several existing papers, which consider Demand Response systems\cite{DR}. However, few of them use reinforcement learning. Reinforcement learning has appeared in numerous books such as "Handbook of Brain Theory and Neural Networks" by Barto\cite{D}, and  "Introduction to Reinforcement Learning" by Sutton and Barto\cite{B}, which we have used because it gives detailed information on the subject and the algorithms used. The main idea behind reinforcement learning and more specifically the Q-learning algorithm is described in the next section.

\section{Reinforcement learning: Q-learning}
Reinforcement learning is a method used to solve problems, formulated as Markov Decision Processes (MDP). It consists of performing a set of experiences which result in positive or negative \textit{rewards} in order to optimize an \textit{objective} function. 
\indent In these types of problems, an active decision-making agent is confronted with an environment, in which it aims to achieve a goal, despite its lack of knowledge of the environment.. The agent's \textit{actions} affect the future state of the environment, so the purpose is to make the agent develop a \textit{policy}, which would point him to the action, giving the best reward at each particular stage.\\
\indent Q-learning, a reinforcement learning technique was first introduced by Watkins in 1989 \cite{W}. The algorithm makes use of a value iteration update. The value, called a Q-function value, is the expected utility of making a particular decision in a particular state. Thus, Q is a function over pairs of states and actions of the agent, who takes the decisions. The updating of a Q-value is performed by taking the old value and making a correction based on the new information learned. The algorithm is proved to be convergent under certain conditions and gives optimal policy for finite MDP.\\
\indent Let's denote with \textbf{s} - a state of the system(\textbf{s'} for the state following \textbf{s}) and with \textbf{a} - an action, which the agent could do. For every state, there exists a set of possible actions. Below, the actions \textbf{a} and \textbf{a'} belongs to the possible sets of actions for states \textbf{s} and \textbf{s'}, respectively. At each step, the agent choose an action for which the Q-value, Q(s,a), is maximized(the agent follows a decision policy derived from Q). The algorithm is as follows:
\begin{algorithm}[H]
\caption{Q-learning algorithm (Sutton and Barto 1998)}
\begin{algorithmic} 
\STATE Initialize \textbf{Q(s,a)} arbitrary, for each addmisible state-action pair (s,a);
\STATE For each episode:
\STATE \hspace{0.5cm} Initialize \textbf{s} with some state of the system.
\STATE \hspace{0.5cm} For each step of episode:
\STATE \hspace{1cm} $\cdot$ Choose an action \textbf{a} from the possible actions in the state \textbf{s}, using policy derived from \textbf{Q}
\STATE \hspace{1cm} $\cdot$ Take action \textbf{a}, observe the reward \textbf{r} and the next state \textbf{s'}
\STATE \hspace{1cm} $\cdot$ Update the Q-value: 
\begin {equation*}Q(s,a) \longleftarrow Q(s,a)+ \alpha[r +\gamma \max\limits_{a' }Q(s',a')-Q(s,a)]
\end{equation*}

\STATE \hspace{1cm} s$\longleftarrow$ s'
\end{algorithmic}
\end{algorithm} 

Here, $\alpha$ and $\gamma$ are parameters, which reflects the learning rate of the algorithm. In our particular case, one day will corresponds to one episode and an hour to a step of episode. The reward $r$ will be the sum that the station earns at the current hour of the day. 

\section{Description of the model}
We are going to describe the details in the station's state representation, as well as the Q-Learning based application which controls the supply decisions of our EVs station, connected to a renewable source of energy.

\subsection{Some preliminary assumptions}
\renewcommand{\labelitemi}{$\bullet$}
\begin{itemize}
\item Discrete time $t_1 = 1,2..,T$. We take T=24. The vehicles arrive dynamically to the station and every arriving vehicle can start the recharging process as early as the beginning of the next hour interval. The station is observed over several number of days.\\
\item  The station has $k$ slots and a maximum number of $M$ participating vehicles (charging or waiting to be charged at any given moment). We can thing that the station has $M$ parking places in total, and at every single time maximum $k$ of the vehicles on them can be supplied.\\

\item When an EV arrives at the station, its driver, which represent the client, states his type which is a function of the vehicle's state of charge (SOC). This function gives the amount of money that the client is ready to pay if his car's battery reaches the different states of charge (for example 5 Euros for SOC=10\% and 15 Euros for SOC=20\%). Knowing a client's type $f(x)$ and his vehicle's  initial SOC $x_{1}$, the intermediate price he is going to pay, in order for his car to have $SOC=x_{2}$ can be easily calculated: $f(x_{2})-f(x_{1})$. For examples of client's type function, see Figure 2 in the next section.\\

\item  Other information that the station receive is the Time To Leave (TTL) of the coming vehicle, i.e. after how many hours the client will expect his vehicle to be ready. \\

\item Additionally, we assume that a renewable source of energy is connected to the station, which supplies the station with variable amount of energy $r(t)$ at each particular $t$ (it is unknown before the moment $t$). The transportation costs are neglected (the cost of transporting the energy from the renewable source to the station slots). Moreover, the station can buy an additional amount of electricity from the grid at any specific time at price $p(t)$ which also varies.
\begin{figure}[H]
\centering
\includegraphics[scale=0.6]{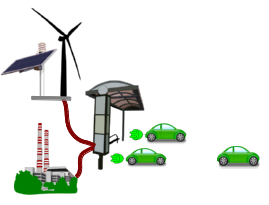}
\caption{An EV charging station connected to the grid and to a renewable energy source}
\end{figure}
\end{itemize}

\subsection{The model}

\indent SAEDS is going to learn the typical total energy demand of the station's customers in the different hours of the day.\\
\indent This approach takes advantage of the fact that the number of clients and their energy consumption at a particular time of the day would be similar each day (weekdays and weekends are not considered separately).\\
\indent The learning approach has to be applied with the help of a big Markov chain with state:
 \begin{equation}\mathbf{\Omega (t) = [t_{1}, USERS(t), r(t), p(t)]}, \end{equation} where:
\begin{enumerate}
\item $t_{1} \in [1..24]$  is the hour of the current day. $t$ is the total number of hours elapsed from the beginning of the first day of the learning phase. In fact, $t_{1}$ is the remainder of $t$, when divided to 24.
\item The data structure USERS(t) consists of 3 column vectors of size $M$- TTL(t), SOC(t) and Types(t).
\\ - The times to leave are integers between 1 and $ttl_{max}$ (the upper limit which we have set is 12).
\\ - SOCs are measured in percents of the full capacity of the battery. We took 10 possible levels- 0,10,20,..100. We assume that every vehicle uses battery with the same capacity.
\\- For Types, we have taken fixed number of functions, chosen in advance. The particular functions that we have used are described in the next section.
\item $r(t)$ is the amount of renewable energy at time $t$. It is also discretized. 
\item  $p(t)$ is the price per unit of extra energy that the station can buy from the grid at time $t$.

\end{enumerate}

\indent We have a queue of customer’s vehicles that is updated dynamically with time.The occurrence of EV arrivals is  modeled with \emph{non-homogeneous Poisson process} i.e.  the number of arrivals in the time interval (a, b], given as N(b)-N(a), follows a Poisson distribution with associated parameter $\lambda_{a,b}$:
\begin{equation}
P[(N(b)-N(a))=k]\\
= \frac{e^{-\lambda_{a,b}}.(\lambda_{a,b})^{k}}{k!},  k=0, .., n
\end{equation}

\indent Thus, the number of arriving customers $z(t)\sim Po(\lambda_{t,t+1})$. Here, the rate $\lambda_{t,t+1}$, $t \in {1,2,...23}$ which is the average number of arriving vehicles during the interval of time [t,t+1], will have to be estimated by monitoring station data from the past. $TTL_{i}(t)$, $SOC_{i}(t)$ and $Types_{i}(t)$ , i=1,2,..,M represent the characteristics of the i-th vehicle in the queue at time $t$.\\
\indent At any given time $t$, the vectors TTL, SOC and Types must be updated as follows: First, when new vehicles arrive, if there are vacant spots among the M places in the station, priority is given on a first-come, first-served basis. After that the online application takes an action. The space of the application's possible actions at time $t$ is formed from all of the vectors $u(t)$, where $u_{i}(t)=0,10$ or $100-SOC_{i}(t)$. The station has $k$ slots in total, so each possible action vector $u(t)$, must have no more than $k$ nonzero components (10 or $100-SOC_{i}(t)$). In other words, at any particular hour, the station can gives the energy equivalence of 10\% SOC or the maximally possible energy equivalence (this, complementing to 100\%) to $k$ or less electric vehicles. This means that our station allows two speed levels of charging: normal charging, which succeed to charge 10\% in one hour, and fast charging, which can charge fully any vehicle, again in one hour.\\
\indent The SOC vector must be updated as follows:
\begin{equation}
SOC(t)=SOC(t)+u(t)
\end{equation}

\begin{equation}
SOC(t+1)=100* \{ SOC(t)/100 \}
\end{equation}
\indent In the latter equation, \{\} means fractional part. In fact, with this equation we nullify the SOCs of the charged vehicles. Times to leave decreases by 1, as we go from time $t$ to $t+1$:
\begin{equation}
TTL(t+1)=TTL(t)-\vec{1}
\end{equation}
\indent Finally, the vehicles, whose TTLs have become 0, must be removed.\\ 
\indent Additionally, at each point in time the array of users is sorted by TTL and if ties occur, by type. In other words, we assume that the array $TTL(t)=[ttl_{1}(t), ttl_{2}(t)...,ttl_{M}(t)]$ has non-decreasing elements for every t and for every sequence of ties $TTL_{i}(t)=TTL_{i+1}(t)=.....=TTL_{j}(t)$ we have $Type_{i}(t)\leq Type_{i+1}(t) \leq.....\leq Type_{j}(t)$. This detail gradually decreases the number of possible states of the system.\\
\indent In order to generate the initial SOCs of the newly arriving EVs, we use manually setted distribution, chosen according to the history of our station. The times to leave (TTL) are dependent on $t$ (for example when an EV comes at 17h, the probability of small deadline before the start of the evening at 20h will increase). For generation of TTLs of the newly arrived EVs, we use normal distribution with different parameters for the different day hours.\\
\indent For $r(t)$ and $p(t)$, real data is used, whose origin is described in the data description section.
\par  The objective function (the station's reward at any given moment) will be:
\begin{equation*}
\Phi (\Omega(t),u(t)) = incomes - expenses
\end{equation*}
\begin{equation*}
incomes = \sum_{m \in M} [Type_{m} (t).(SOC_{m}(t)+u_{m}(t)) - Type_{m} (t).SOC_{m}(t)]
\end{equation*}
\begin{equation*}
expenses = p(t).[(\sum_{m \in M} u_{m}(t))  - r(t)] 
\end{equation*}

\indent We have a value function over states  $V(\Omega)$  which represents the “profitability” of the state $\Omega$.
\begin{equation}
V(\Omega)= \min\limits_{u} (\Phi (\Omega , u) + E[V(\Omega_{1})])
\end{equation}
In the latter equation, $(\Omega_{1}$ is the station's state, following $(\Omega$.
\indent Q-learning is an on-line, reinforcement learning method, which approximates the value of the function $V (\Omega)$. Q-learning estimates
the value of a Q-function for each state-action pair $(\Omega, u)$ and 
\begin{equation}
V(\Omega)= \min\limits_{u} Q(\Omega , u)
\end{equation}
\par In summary, the Markov chain will have states in the form of $\Omega$. The transitional probabilities depend on $\lbrace r(t), p(t)\rbrace$ and are unknown. Let us describe the essence of the Q-learning algorithm:
\renewcommand{\labelitemi}{$\bullet$}
\begin{itemize}
\item At k = 0, initialize $Q_{k}(\Omega, u)$  for every state-action pair $(\Omega, u)$
and select initial state $\Omega.$
\item Choose $u = \argmin Q_{k} (\Omega, u) $ with probability $1 - \alpha_{t},$
else let u be a random exploratory action(uniformly selected).
\item Carry out action u. Let the next state be $\Omega '$, and the cost
be $\Phi (\Omega , u)$. Update the Q-values as follows:
\begin{multline}
Q_{k+1}(\Omega , u)= (1-\beta_{k}).Q_{k}(\Omega , u) \\ + \beta_{k} .(\Phi (\Omega , u) + \min\limits_{u} Q(\Omega ' , u))
\end{multline}
\\ Here, $\beta_{k}$ and $\alpha_{t}$ are some probabilities chosen in advanced.
\item Set the current state to $\Omega '$, increment k and go to step
2.
\end{itemize}
\section{Computer simulation}
\subsection{Users' type function}
Users' types are defined by customer's willingness to pay in order to have their vehicle charged if their initial SOC is 0\%. The price that the user pay is the difference between the price of the SOC they want to reach and their initial SOC. In the simulation, we consider two types of users: the so called \textit{rich} type of users and \textit{medium} type of users. We model their types' function by an utility function represented by the quadratic function  \eqref{eq:test}. This 
family of functions was choosen, because users pay less money for each extra 10\% level of charging. In fact, this functions belongs to the family of sigmoid functions, which are widely used in practice.
\begin{equation} \label{eq:test}
f(x)=\left\{
	\begin{array}{ll}
		\dfrac{max}{100}(2x-\frac{1}{100}x^{2})  & \mbox{if } 0 \leq x \leq 100 \\
		max & \mbox{if } x > 100
	\end{array}
\right.
\end{equation}
\indent The variable $max$ refers to the price of fully charging an initially empty vehicle. Example curves are shown in the figure below, where $max_{rich} =3.6$ and $max_{medium} =2.4$
\begin{figure}[H]
\centering
\includegraphics[scale=0.7]{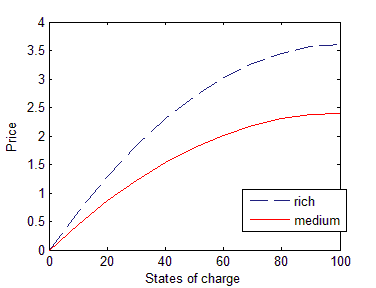}
\caption{Users' types}
\end{figure}
\subsection{Data description}
Two categories of data were used - renewable energy data and pricing data. \\ 

2.1 Renewable energy  \\
\par As explained before, we suppose that the station is connected to a wind energy generator (wind turbine for example) and has a solar panel.\\
\indent We need data about the average wind generation from a wind energy generator at hourly slots of the day. This data can be found on \textit{R\'eseau de transport d'\'electricit\'e} (RTE) France website \cite{E} and is provided as an Excel file. The data is about wind generation across France. For the sake of consistency, we divided all the numbers by 4058 (the number of wind turbines in France in 2012)\cite{F}.\\
\indent For the solar panel generation, we assume that the EV charging station has a 10 $m^{2}$ solar panel which is capable of generating an average of 1000 kWh per year in France\cite{F}. In order to calculate the amount of electricity per hour, we assume that  solar panel generation follows a normal distribution. The generation peak is assumed to occur at 1.30 p.m. \\ 

2.2 Pricing \\
\par The RTE France website also provides data about electricity prices at half hourly slots. We take into account this data in our model by taking the average price at every hour.
\subsection{Simulation parameters and  results}
A comparison between the income (in euro) of a small EV station using SAEDS and the income of a station making random decisions at each point in time is depicted on the graph below. The income was measured for 29 consecutive days after a training period with data for 190 days, repeated 40 times. The results shows that  the income increases by 81 percents, which is a good result, in spite of the fact that the purely random strategy can be easily beaten with the set problem formulation. However, such a percentage is an evidence for a high learning effect. As a result of previous executions of the program, we have received slightly lower results, in the range of 40 to 80 percents of income increasing.
\par Parameter values that were used to obtain these results are:
 \\ $M=5$, $k=3$ and  the 2-type functions already described(rich and medium). For r(t) and p(t)- for simplicity, we took only two possible values- high and low. For $\alpha$ and $\beta$(parameters of the learning process) were taken linear by $t$ functions.
\begin{figure}[H]
\centering
\includegraphics[scale=0.3]{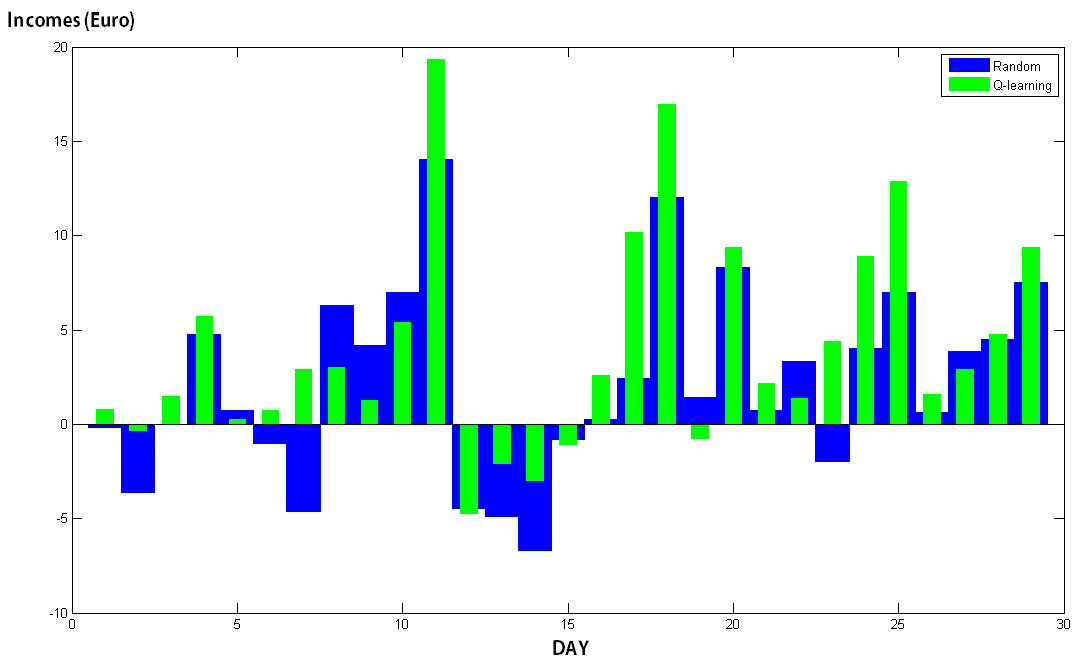}
\caption{Income (in euro) comparison between an EV station which uses our algorithm (green), and an EV station that uses random decisions (blue). }
\end{figure}
\section{Conclusion}
We tried to apply an already existing approach \cite{A} of learning an optimal decision policy, but in the context of electric vehicles charging. There are several similarities between the problem, formulated in \cite{A} and the study formulated in our paper, which portended good final results. After implementing the model on a computer, it was observed that our algorithm outperformed the most trivial one (uniform distribution over each possible decision, on every step). The increase of the station's incomes in the range 40-80\% shows that using this learning approach is meaningful.
\par We can conclude that the proposed learning scheme performs at a satisfying level. Nevertheless, the model can be adjusted in order to mimic reality more closely. Probably, some of the numerical parameters in the algorithm can be adjusted more precisely, in order to receive bigger increase of the incomes. However, the utility of the proposed learning scheme is obvious.

\ifCLASSOPTIONcaptionsoff
  \newpage
\fi

\end{document}